\documentstyle[preprint,prb,aps]{revtex}

\begin{document}

\title{Statistics of radiation in a disordered medium:\\
from ballistic to diffusive regime}

\author{Eugene Kogan and Moshe Kaveh}

\address{Jack and Pearl Resnick Institute of Advanced Technology, \\
Department of Physics, Bar-Ilan University, Ramat-Gan 52900, Israel.}

\maketitle

\begin{abstract}
The distribution function for the intensity of radiation propagating in a
random medium is analyzed for arbitrary multiplicity of scattering (for
arbitrary relation between the distance of propagation and mean free path),
including as limiting cases purely ballistic and purely diffusive regimes.
The calculation is fulfilled in the framework of diagram techniques and is
based on the topological analysis of diagram.
\end{abstract}
\bigskip

When electromagnetic wave propagates in a random medium and undergoes
multiple scattering the density of energy in a given point (intensity) $%
I=<E^2>$ is a strongly fluctuating quantity. In a diffusive regime these
fluctuations are traditionally described by Rayleigh statistics \cite{good}.
The distribution function $P_R(I)$ is a simple negative exponential:
\begin{equation}
\label{1}P_R(I)={\frac 1{<I>}}\exp \left( -{\frac I{<I>}}\right) ,
\end{equation}
which corresponds to the equation for the moments:
\begin{equation}
\label{2}<I^n>=n!<I>^n.
\end{equation}
In Eqs. (\ref{1}) and (\ref{2}) the averaging is with respect to all
macroscopically equivalent but microscopically different scatterers
configurations. By diffusive we mean the regime when it is possible to
discard the ''coherent'' component of the field, that is

\begin{equation}
\label{xx}<E>=0.
\end{equation}
This assumption is true for the case $r/\ell \gg 1$, where $r$ is a distance
from the source to the observation point, and $\ell $ is a mean free path
(for the sake of definiteness we consider in this paper only the case of
point source in an infinite medium). The aim of the present publication is
to get the distribution function for arbitrary relation between $r$ and $%
\ell $, when generally speaking the ''coherent'' component of the field
should also be taken into account.

Here some clarification should be made. In our previous publications \cite
{spec1} we presented systematic derivation of the intensity distribution
function in a diffusive regime and have shown that the Rayleigh statistics
is correct only for $I/<I>\ll k\ell $, where $k$ is the radiation wave
vector. For larger values of intensity the distribution function differs
drastically from a simple exponential and the asymptotic behavior is a
stretched exponential. In this paper we consider the case of $k\ell $ is so
large that this effect can be ignored (or alternatively we are not
interested in the distribution function tail). But if in the above mentioned
publications we restricted ourselves by the case $r/\ell \gg 1$, now we are
interested in the case of the arbitrary value of the parameter $r/\ell $ .

To achieve this aim we apply the same kind of reasoning based on the
topological analysis of the diagrams we used previously \cite{spec1}. To get
the idea of it let us remember how Rayleigh statistics is obtained in the
framework of the traditional diagram technique \cite{shap}. The point source
is situated at the origin; the intensity is measured in the point $r$. In
the diagrammatic representation $<I>$ is given by the diagrams with a pair
of wave propagators $G_{0r}^R$ and $G_{0r}^A$, summed with respect to all
possible interactions with the scatterers. The n-th moment $<I^n>$ is given
by the set of diagrams with n propagators $G_{0r}^R$ and n propagators $%
G_{0r}^A$. For us is important the following property of diagrams: if the
diagram consists from several disconnected parts then the contribution of
that diagram is equal to the product of contributions of disconnected parts.
It means, that if we consider for $<I^n>$ only the diagrams consisting of $n$
disconnected parts, each part being the set of diagrams with a pair of
propagators (advanced and retarded), we immediately obtain Eq.~(\ref{2}).
The multiplier $n!$ which appears in the n-th moment is of combinatorial
origin; it is simply the number of possible pairings between propagators.

So to get Rayleigh statistics we have ignored the connected diagrams with
two or more pairs of propagators. In Ref. \onlinecite{spec1} we have formulated
a
perturbation theory which systematically takes such diagrams into account,
ant it is such diagrams which lead to deviations from negative exponential
statistics for large values of intensity. But we also ignored the diagrams
which contain isolated (dressed) propagators. The ground for this is that
each isolated propagator gives contribution of the order of $\exp (-r/\ell )$%
, and in the diffusive regime one can ignore such diagrams. Here we should
take them into account. That is for $<I^n>$ we consider only diagrams which
consists of isolated propagators and pairs of propagators (advanced and
retarded one). The sum of all diagrams can be written down in the following
way:
\begin{equation}
\label{3}<I^n>=\sum_{m=0}^nP(n,m)A^mB^{n-m},
\end{equation}
where $A=\left| <E>\right| ^2$, $B=<I>-|<E>|^2$, and $P(n,m)$ is the
coefficient of purely combinatorial origin:
\begin{equation}
\label{4}P(n,m)={\frac{(n!)^2}{(m!)^2(n-m)!}.}
\end{equation}

The distribution function $P(I)$ is connected with its moments by usual way:
\begin{equation}
\label{6}P(I)=\int_{-\infty }^\infty \exp (i\xi I)~\sum_{n=0}^\infty {\frac{%
(-i\xi )^n}{n!}}<I^n>~{\frac{d\xi }{2\pi }}.
\end{equation}
Substituting the Eq. (\ref{3}) into the Eq. (\ref{6}) and changing the order
of summation,which is possible due to convergence of the series, we get:

\begin{equation}
\label{5a}P(I)=\int_{-\infty }^\infty \exp (i\xi I)~\sum_{m=0}^\infty {\frac{%
\left( -i\xi A\right) ^m}{(m!)^2}}\sum_{n=m}^\infty {\frac{n!}{{(n-m)!}}%
\left( -i\xi B\right) ^{n-m}}~{\frac{d\xi }{2\pi }.}
\end{equation}
Taking into account that for $n\geq m$

\begin{equation}
\label{5b}{\frac{n!}{{(n-m)!}}X^{n-m}=\ \ \frac{d^m}{dX^m}{X}^n,}
\end{equation}
we obtain:%
$$
{}~\sum_{m=0}^\infty {\frac{\left( -i\xi A\right) ^m}{(m!)^2}}%
\sum_{n=m}^\infty {\frac{n!}{{(n-m)!}}\left( -i\xi B\right) ^{n-m}}~=~\frac
1{1+{i\xi B}}\sum_{m=0}^\infty {\frac 1{m!}\left( \frac{{-i\xi A}}{1+{i\xi B}%
}\right) ^m}
$$

\begin{equation}
\label{5c}{{=\frac 1{1+{i\xi B}}\exp {\left( \frac{{-i\xi A}}{1+{i\xi B}}%
\right) }}.}
\end{equation}
So finally Eq. (\ref{6}) can be written in the form:

\begin{equation}
\label{9}P(I)=\int_{-\infty }^\infty {\frac{\exp \left( {i\xi I-{\frac{i\xi a
}{1+i\xi (1-a)}}}\right) }{1+i\xi (1-a)}}~~{\frac{d\xi }{2\pi }}
\end{equation}
(the intensity is measured in the units of $<I>$ and  parameter $%
a=A/(A+B)=|<E>|^2/<I>$,  is introduced). It is obvious that for $a=0$
(purely diffusive transport), the
Eq. (\ref{9}) coincides with the Eq. (\ref{1}), so this parameter gives the
degree of deviation from Rayleigh statistics. In the opposite limiting case $%
a=1$ the Eq. (\ref{9}) gives
$P\left( I\right) =\delta \left( I-1\right) $.
This limiting case means that there are no fluctuations,
which can occur only when there is no scattering at all, in other
words in the ballistic regime.

For intermediate
value of $a$ the distribution function can easily be calculated
numerically.  This behavior is represented  on Fig. 1. We see that even for
 $a\ll 1$ (almost diffusive transport)  taking into account of the ''coherent''
component of the  field drastically changes the distribution function
 for small $I$; in
particular for $I=0$  we get: $P\cong 1/2$. Also interesting are the predicted
oscillations of the distribution function.\

\acknowledgements The authors acknowledge the financial support of the
Israeli Academy of Sciences.\

\begin{figure}
\caption{ Distribution function for: (1) $a = .1$ (solid line), (2) $a = .5$
(dashed line),
(3) $a = .9$ (dot-dashed line).
\label{fig1}}
\end{figure}


\begin{thebibliography}{9}
\bibitem{good}  J. W. Goodman, {\it Statistical Optics} (J. Wiley, New York,
NY, 1985).

\bibitem{spec1}  E. Kogan, M. Kaveh, R. Baumgartner, and R. Berkovits, Phys.
Rev. B {\bf 48}, 9404 (1993); Physica {\bf A 200}, 469 (1993).

\bibitem{shap}  B.Shapiro, Phys. Rev. Lett, {\bf 57}, 2168 (1986).
\end{thebibliography}
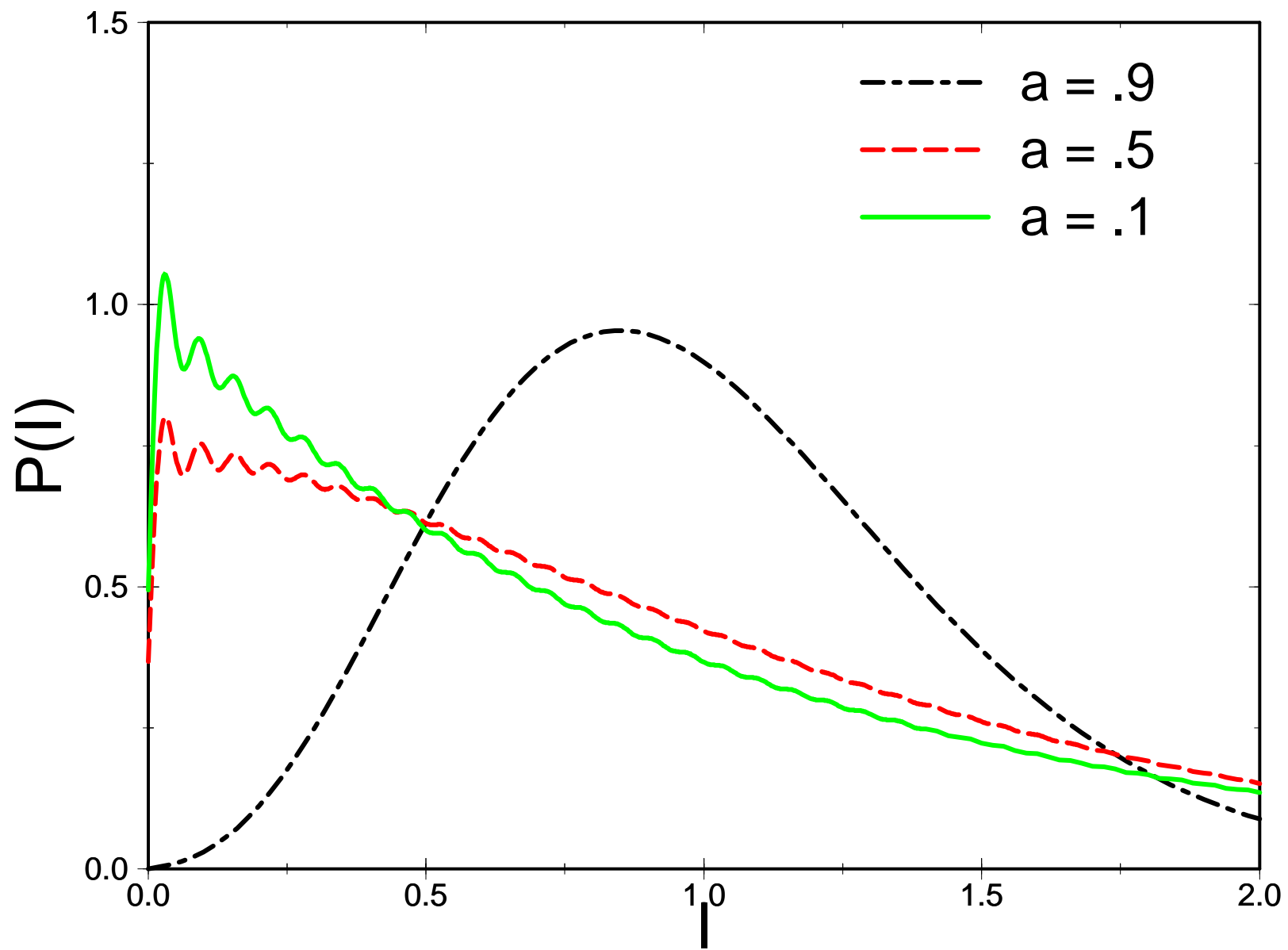
\end{document}